# CIPARSim: Cache Intersection Property Assisted Rapid Single-pass FIFO Cache Simulation Technique

Mohammad Shihabul Haque    Jorgen Peddersen    Sri Parameswaran
School of Computer Science and Engineering
University of New South Wales, Australia
Email: {mhaque,jorgenp,sridevan}@cse.unsw.edu.au

*Abstract*—An application's cache miss rate is used in timing analysis, system performance prediction and in deciding the best cache memory for an embedded system to meet tighter constraints. Single-pass simulation allows a designer to find the number of cache misses quickly and accurately on various cache memories. Such single-pass simulation systems have previously relied heavily on cache inclusion properties, which allowed rapid simulation of cache configurations for different applications. Thus far the only inclusion properties discovered were applicable to the Least Recently Used (LRU) replacement policy based caches. However, LRU based caches are rarely implemented in real life due to their circuit complexity at larger cache associativities. Embedded processors typically use a FIFO replacement policy in their caches instead, for which there are no full inclusion properties to exploit. In this paper, for the first time, we introduce a cache property called the "Intersection Property" that helps to reduce single-pass simulation time in a manner similar to inclusion property. An intersection property defines conditions that if met, prove a particular element exists in larger caches, thus avoiding further search time. We have discussed three such intersection properties for caches using the FIFO replacement policy in this paper. A rapid single-pass FIFO cache simulator "CIPARSim" has also been proposed. CIPARSim is the first single-pass simulator dependent on the FIFO cache properties to reduce simulation time significantly. CIPARSim's simulation time was up to 5 times faster (on average 3 times faster) compared to the state of the art single-pass FIFO cache simulator for the cache configurations tested. CIPARSim produces the cache hit and miss rates of an application accurately on various cache configurations. During simulation, CIPARSim's intersection properties alone predict up to 90% (on average 65%) of the total hits, reducing simulation time immensely.

## I. INTRODUCTION

In a computer system, energy consumption, execution time and overall system performance during execution of an application are greatly influenced by both the cache miss rate and the configurations (combinations of different cache parameters such as the number of cache sets (set size), associativity, line size (block size), etc.) of the cache memories in the memory hierarchy. Cache miss rates of the same trace using various cache configurations are generally unpredictable. Hence we need to find the cache miss rates in different cache configurations to decide the most suitable configuration. When the total number of cache misses is known for a particular application and cache memory, using analytical models such as the one proposed in [19], energy consumption by the subject cache memory, execution time for the application and overall system performance can be estimated quickly. Therefore, in deciding the best cache memory for an embedded system [14] [13] [19] [27] and in timing analysis [26] [28], cache miss rates of applications are widely used. Thus, to determine the best cache configuration, given power area and performance constraints, cache miss rates for all configurations must be found first.

To save time in detecting the cache miss rate of an application on a particular cache memory, simulation of the application's memory access trace with the least possible hardware details is widely used instead of real application executing on real cache memory. A robust, time saving and resource generous variant of the trace driven simulator is the single-pass simulator (e.g., [12] [13] [14] [29]). In a single-pass simulator, multiple cache configurations are simulated together while reading one application's trace of memory accesses only once. Besides reducing trace reading time, single-pass simulators deploy several other speedup mechanisms such as customized data structures to represent cache memories and to search and update data in the cache memories quickly (e.g., associativity list in [19], "Wave" in [13], "CLT" in [14], etc.). In addition to custom-tailored data structures, use of cache inclusion properties, introduced by Mattson et al. in [22], is also popular in reducing cache simulation time. An inclusion property indicates when all the elements within one cache configuration are known to be present in other configurations. Therefore, inclusion properties allow some of the simulation steps to be avoided, saving simulation time enormously when a large group of cache configurations are simulated together.

Cache inclusion properties do not hold for First-In-First-Out (FIFO) caches [22]. Previous studies have predicted the status of single FIFO cache configurations [9] [10] [25] [26]. However, the methods in those articles do not translate well to predict the contents of multiple caches in single-pass simulation simultaneously.

As a cache replacement policy, FIFO has several advantages. Among the replacement policies, caches with FIFO replacement policy demonstrate lower energy consumption, especially compared to a cache with LRU replacement policy [4]. Their simple design makes FIFO caches inexpensive to implement. Due to these reasons, FIFO is widely used as the cache replacement policy in embedded processors (e.g., Tensilica Xtensa LX2 processors [32], Intel XScale [1], ARM9 [3] and ARM11 processors [2]). Therefore, a fast simulator to decide the cache miss rate of an application on various FIFO caches is indeed in a great demand. To meet this demand, smart data structure based cache simulators such as [13] [14] are in use. However, the possibility to utilize cache inclusion properties in addition to smart data structures would be of great use in reducing simulation time further. To the best our knowledge, no simulator has ever been proposed that utilizes any FIFO cache property to reduce single-pass simulation time significantly.

In this paper, for the first time, we take an initiative to reduce single-pass simulation time for FIFO caches utilizing some FIFO cache properties. We introduce a cache property, the "Intersection property", that can help to speed up cache simulation using the same principle of inclusion properties. We have presented three cache intersection properties for the FIFO replacement policy. Utilizing these three intersection properties and custom tailored space and time saving data structures, we have proposed a new single-pass FIFO cache simulator "CIPARSim". CIPARSim is the first of its kind to utilize any FIFO cache property, such as the intersection property, to speedup simulation. Experimental results show that CIPARSim outperforms the available single-pass FIFO cache simulators SCUD [14] and DEW [13] significantly for all the SPEC

CPU2000 [16] and Mediabench [20] applications tested. We consider SCUD as the state of the art single-pass FIFO cache simulator as DEW can simulate only caches with varying set sizes in a single pass over the trace file.

**Problem statement:** Given an application's memory access trace and a set of cache configurations using the FIFO replacement policy, reduce the simulation time to find the cache miss rates of all given cache configurations executing the trace by utilizing FIFO cache properties.

**Layout:** The rest of the paper is structured as follows. Section II presents the related works, Section III introduces the concept of cache intersection properties and presents three FIFO cache intersection properties, Section IV describes the new rapid single-pass FIFO cache simulator CIPARSim with its custom tailored data structures, Section V describes the experimental setup and discusses the results found for SPEC CPU2000 and Mediabench applications; and Section VI concludes the paper.

## II. RELATED WORK

Mechanisms for acceleration of trace driven simulation to find cache miss rate have been studied for a long time for further improvement. Depending on the accuracy of the simulation results, these acceleration techniques can be categorized into two categories. The methods with limited accuracy are called estimation methods (E.g., [7] [18], etc.). These heuristics dependent methods are fast; however, not preferred when accuracy of simulation result is required. Several proposals for acceleration of trace driven cache simulation without affecting the accuracy of the results have been proposed, too. These proposals can be broadly categorized into (i) Compressed trace simulation, (ii) Parallel simulation and (iii) Single-pass simulation.

In a compressed trace simulation, redundant information are pruned to compress the application's memory access trace. As the compressed trace is often considerably shorter than the actual memory access trace, simulation time can be reduced significantly. However, success of these simulators relies on the compressibility. In addition, time to compress and decompress the trace file accurately adds overhead to the actual cache simulation time. Some examples of compressed trace simulation approaches are [23] [30] [31].

To reduce the overall simulation time, several proposals, called "Parallel simulation", were made to perform the simulation of a group of cache configurations in parallel on multiple processors. Depending on the source of parallelism, these proposals can be categorized into several subcategories. The proposal in [5] is based on set-parallelism and simulates each cache set of a cache configuration on different processors. Han et al. proposed a method in [11] that not only exploits set-parallelism but also parallelizes searches for the requested data block in a particular cache set. Similarly, Heidelberger et al. introduced time-parallelism in [15] and Nicol et al. proposed stack distance based parallel simulation in [24]. Parallel simulation methods undoubtedly speed up the simulation process. However, their main limitation is in the high resource demand to perform simulations in parallel. Due to their resource hungry behavior, implementation is costly too.

In contrast to parallel simulation, one processing unit is used as optimally as possible in a single-pass simulation approach. Therefore, single-pass simulation can be combined with parallel or compressed trace simulation for further speedup. Single-pass simulation approaches usually exploit inclusion properties and custom-tailored data structures to reduce processing time without any help of extra hardware. In the article [17], published in 1989, Hill et al. studied the effect of varying associativity in caches in search for a rapid single-pass cache simulation approach. Sugumar et al. [27] made an effort to exploit the cache inclusion properties when they introduced the use of binomial trees to speed up LRU single-pass cache simulation in 1995. Their proposed method improved the method of [17]. Utilizing binary trees and some cache inclusion properties based on the LRU replacement policy, Sugumar's method was able to simulate multiple cache configurations very quickly in a single pass over an application trace. In 2004, Li et al. [21] proposed an advancement to Sugumar's proposal through a compression method to reduce simulation time. In 2006, Janapsatya et al. [19] proposed a method to traverse the binary tree in a top-down fashion to exploit the temporal locality in cache line accesses. In 2009, the CRCB algorithm [29] improved the simulation time of Janapsatya's technique by using a runtime pruning of the trace file and two inclusion properties. In 2009, Haque et al. [12] showed that, instead of top-down traversal, bottom-top traversal of the binary simulation tree will enable the simulator to exploit a different set of inclusion properties for LRU caches. In 2010, DEW [13] and SCUD [14] were proposed to perform rapid single-pass simulation of FIFO caches exploiting custom-tailored data structures. However, the space hungry behavior and time consuming manipulation of those custom tailored data structures left room for further improvement. To the best of our knowledge, until today no cache property has been proposed for FIFO caches to be exploited in the single-pass simulation for acceleration of operation.

### A. Our contributions

1) For the first time, a cache property called the "Intersection property" has been introduced in this article to predict the existence of a memory block content in multiple FIFO caches during single-pass simulation.

2) Three FIFO cache intersection properties have been proposed that can be used to reduce simulation time in FIFO single-pass simulation.

3) A rapid single-pass FIFO cache simulator "CIPARSim" has been proposed that, utilizing the proposed FIFO intersection properties, shows significantly faster performance than the available FIFO cache simulators. To the best of our knowledge, CIPARSim is the first single-pass cache simulator that uses FIFO cache properties to reduce simulation time.

4) Experiment results have been presented for SPEC CPU2000 and Mediabench applications to show the effectiveness of the intersection properties and the cache simulator CIPARSim.

## III. INCLUSION PROPERTIES VS. INTERSECTION PROPERTIES

In [22], Mattson et al. defined inclusion properties as a condition where larger caches contain a superset of smaller caches at every point in time. This holds between alternative caches that have the same cache line size, do not pre-fetch, have the same number of sets, and the replacement policy must induce a total priority ordering on all previously referenced memory blocks (that map to each cache set) before each reference and use only this priority ordering to decide the next replacement cache block. The LRU replacement policy is shown to have this feature.

When inclusion properties hold between two caches, just by simulating the smaller cache, we can realize which memory block contents will be available in the larger cache. Therefore, when any of the contents from the smaller cache is re-accessed, simulation can be avoided in the larger cache for that access.

Similar to inclusion properties, intersection properties we propose predict the availability of one or more particular memory block addresses in certain cache configurations based on their presence in a different configuration. However, intersection properties require particular conditions as opposed to being inherent like the inclusion properties. Intersection properties can be used when a cache replacement policy such as FIFO does not show inclusion properties.

Therefore, if caches A and B have the same replacement policy and the caches differed by associativity, set size, cache line size or all, their replacement policy shows intersection property when, on a particular condition, *{Contents of A} ∩ {Contents of B} = ∅* and the contents inside *{Contents of A} ∩ {Contents of B}* are predictable using the condition. An example of a cache intersection property is, "The most recently accessed memory block of a cache memory will be available in any other cache memory as the most recently accessed memory block" (proposed as an inclusion property in the CRCB [29] algorithm). During single-pass simulation, if an intersection property condition is met for one cache configuration, the simulation of the current access can be avoided in the cache configurations predicted by the intersection property. Note that inclusion properties are also intersection properties. However, not all intersection properties are cache inclusion property.

*A. FIFO intersection properties*

We now present and prove several FIFO cache intersection properties useful for the rapid single-pass simulation of FIFO cache configurations. In this context, it is assumed that cache line size is constant for all considered cache configurations. The terms 'larger' or 'smaller' applied to cache configurations refer to the set size and/or associativity. E.g., a 'larger' cache has equal or greater set size and equal or greater associativity, with at least one being greater.

**Intersection Property 1:** *When an element is inserted into a FIFO cache of associativity $A_X$, if the element's location within a large FIFO cache with associativity $A_Y$ is at least $((2 \times A_X) - 3)$ elements away from the replacement pointer of the larger cache in the direction of replacement, it guarantees the existence of the memory block content in the larger cache at least as long as it remains in the smaller cache.*

*Proof:* To prove the first cache intersection property, we have to figure out the maximum number of insertions possible in the larger cache, after insertion of an element in the smaller cache. By analysis, the situation that will cause this to occur is as follows:

Let 'I' be the element that is most recently inserted (MRI) into the smaller cache. Let there also be one element in the smaller cache which is missing in the larger cache. We will call this uncommon element 'U'. In order for 'U' to exist, there must be an element 'R' in both caches that replaced 'U' in the large cache. The 'remaining' $(A_X - 3)$ elements in the smaller cache appear at and following the replacement pointer in the larger cache. Figure 1(a) shows such a layout using 8-way and 16-way caches. The 'remaining' elements are lettered 'A' through 'E' and the other elements in the larger cache are labeled as '-' as their value do not matter. 'LRI' indicates the least recently inserted element that suppose to be replaced at the next insertion.

The access pattern required is to first access element 'U' which replaces the first 'remaining' element ('A' in the example). The 'remaining' elements are then accessed in the order they were in the larger cache as, each will replace the next ('A' replaces 'B', 'B' replaces 'C', etc.). After these, $(A_X - 1)$ new elements can be accessed which will replace all elements except 'I' in the smaller cache as well as $(A_X - 1)$ elements in the larger cache. This is the largest amount of replacements that could occur in the larger cache without replacing 'I' in the smaller cache and the total number of insertions is $1 + (A_X - 3) + (A_X - 1) = (2 \times A_X - 3)$. The final situation in our previous example is shown in Figure 1(b) with the new elements that were inserted marked as '+'. Thus, as long as 'I' is located $(2 \times A_X - 3)$ elements away from the initial replacement pointer (as shown in the example) it can never be removed from the larger cache without removing it from the smaller cache at the same time or earlier.

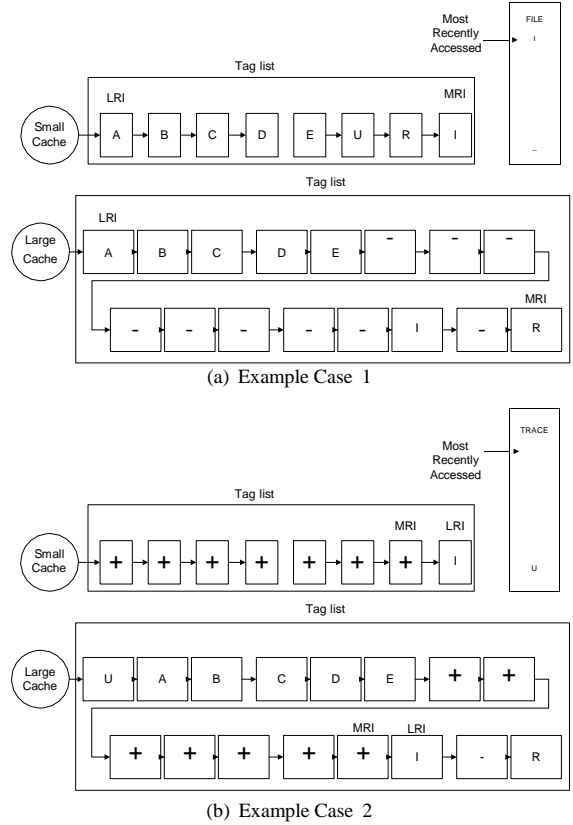

(a) Example Case 1

(b) Example Case 2

Fig. 1. Example Cases for The First Intersection Property

**Intersection Property 2:** *Using the FIFO cache replacement policy, the most recently inserted (MRI) element of any set in a 2-way associative cache (each cache set with two lines) of set size $S_{2-way}$ must be present in all larger FIFO cache configurations, i.e., those with set size $S \geq S_{2-way}$ and associativity $A \geq 2$.*

*Proof:* Call the MRI element of a set within a 2-way cache '$X$' and the other element '$Y$'. Call the time immediately before '$X$' was inserted $t_1$, and the time when '$X$' was inserted $t_2$.

At $t_2$, '$X$' is the most recently accessed element and is thus present in all cache configurations. The counter-example to the intersection property can only occur if '$X$' is not present in the larger cache at a future time $t_3$. Only '$X$' and '$Y$' can be accessed between $t_2$ and $t_3$ as otherwise, '$X$' would no longer be the MRI. The counter-example requires that '$X$' is not present in the larger cache at $t_3$, yet it was present at $t_2$, so it

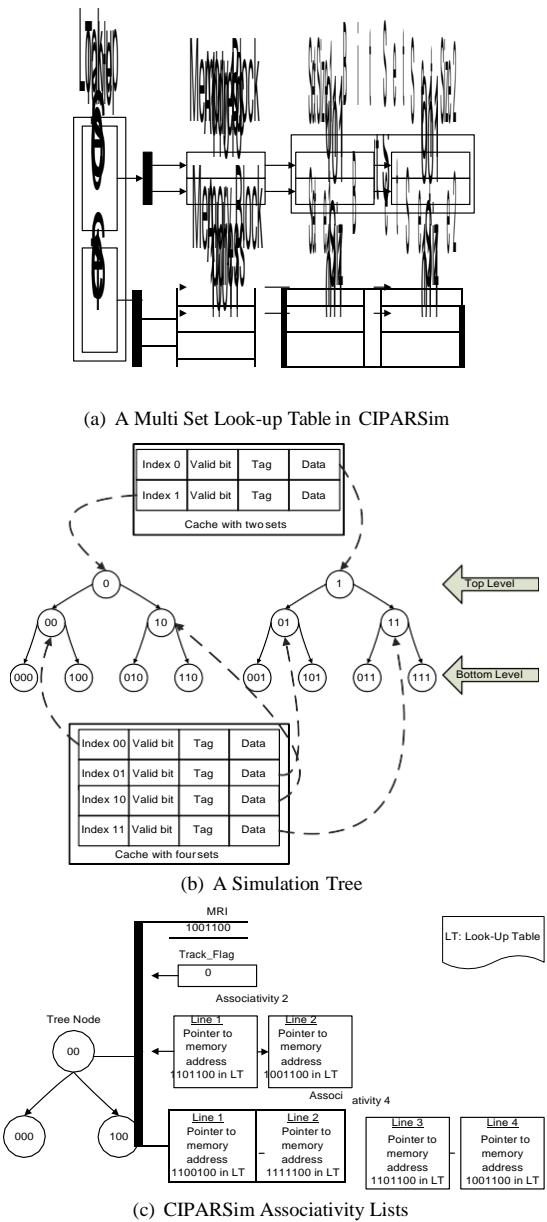

(a) A Multi Set Look-up Table in CIPARSim

(b) A Simulation Tree

(c) CIPARSim Associativity Lists

Fig. 2. CIPARSim Data Structures

```
Function AddressEvaluation(RA)
1  if (RA is not found in LT) then
2      Record one cache miss for all the cache configurations;
3      Place RA in LT and place pointer to RA's location in LT in all the cache
       configurations in the simulation tree;
4      if (the smallest A = 2) then
5          In all the configurations with A = 2, make RA the MRI and set
           Track_Flag=False;
6      else
7          Set the Intersection_Flag=True for RA in the smallest associativity
           configurations;
8  else
9      select the tree level L = 0 (smallest cache set size S = 2^L);
10     while 2^L is not larger than the largest set size do
11         A = the smallest associativity;
12         if (A = 2) then
13             if (RA is found in A) then
14                 if (Track_Flag = True) then
15                     Record cache hit in all the remaining caches;
16                     return AddressEvaluation(RA);
17                 else
18                     if (MRI = RA) then
19                         Set Track_Flag = True;
20                         Record cache hit in all the remaining caches;
21                         return AddressEvaluation(RA);
22                     else
23                         Set Track_Flag = True;
24                         Record cache hit for the selected cache;
25                         A = A × 2;
26             else
27                 Change Track_Flag = False and make RA the MRI
                   of the configuration with A and S = 2^L;
28         else
29             if (RA is found in A) then
30                 if (Intersection_Flag is True) then
31                     Record cache hit for all the caches with the selected
                       cache set size S;
32                     jump to line 45;
33                 else
34                     Record cache hit for the selected cache;
35                     A = A × 2;
36             else
37                 Update Intersection_Flag (see Section IV-A);
38         while A is not larger than the largest associativity do
39             if (RA is not in A) then
40                 Record a cache miss for the selected cache;
41                 Place a pointer to the RA's location in LT in the selected
                   cache;
42             else
43                 Record a cache hit for the selected cache;
44             A = A × 2;
45         L = L + 1;
```

associativity 2, the Track Flag is set to false (or 0). If an existing memory block of the associativity 2 with the selected tree node is re-accessed, the Track Flag is set to true. The Track Flag helps to exploit the *third intersection property* of Section III-A. A cache hit in the FIFO associativity 2 with Track Flag set to true indicates that the memory block content is available in all the larger FIFO cache configurations in the same simulation tree. If the smallest associativity is larger than two, a bit "Intersection Flag" is associated with **each cache line in the smallest associativity list** of a simulation tree node. These extra bits will help to utilize the *first intersection property* of Section III-A. Whenever a new memory block tag is inserted in a cache line in the smallest associativity list of a tree node, the Intersection Flag is set to true if the same memory block tag is at least $((2 \times A_X) - 3)$ (where $A_X$ is the smallest associativity) elements away from the replacement pointers in the other larger associativity lists in the same tree node. Therefore, when a memory block content in the smallest associativity in a tree node is re-accessed, simulation can be avoided in the larger associativities if the Intersection Flag is found true. In CIPARSim, the smallest associativity must be 2 or larger if MRA tag is not save for each simulation tree node separately to simulate direct mapped caches.

In Figure 2(c), an example tree node '00' from Figure 2(b) is presented with two FIFO associativity lists illustrated (namely associativity=2 and associativity=4). Node '00' is from the second level in the tree of Figure 2(b). The second level of the tree of Figure 2(b) represents a FIFO cache with set size 4. The first cache line of the list for associativity 2 has a pointer to the memory block address "1101100" in the look-up table of CIPARSim. Using these pointers, CIPARSim can update the look-up table's bit arrays when the address "1101100" will be evicted from the associativity 2's list in tree node '00' due to a miss for that node.

### B. CIPARSim simulation approach

To simulate an application trace, CIPARSim reads one requested memory block address at a time from the trace file and evaluates it in the FIFO cache configurations under simulation. CIPARSim does not simulate consecutive request for the same address. For a requested memory block address, cache hit/miss evaluation continues from the smallest to the largest FIFO cache set size in the look-up table. For a particular cache set size, cache hit/miss evaluation continues from the smallest to the largest associativity lists before moving to the next set size. In other words, cache simulation starts in the top level's smallest associativity list in a simulation tree and finishes in the bottom level's largest associativity list.

Function AddressEvaluation illustrates the process to evaluate an address request (*RA*) to determine hit and miss for the FIFO cache configurations (with set size $S = 2^L$, associativity *A* and same cache line size) under consideration. We assume that the associativities are $2^i$ where $i \geq 1$ and the smallest set size is 1 to simulate fully associative caches. In the function, *LT* represents Look-up Table. A textual description of the flow of the Function AddressEvaluation is given below:

**1.** *RA* evaluation starts from searching the address in the appropriate *LT* set using **binary search**. If the address is not found in the look-up table, CIPARSim declares a cache miss for all the cache configurations. *RA* is placed in the *LT* and a pointer to *RA*'s location in *LT* is placed in every cache configuration. *RA* is placed in the *MRI* of associativity 2 and associativity 2's *Track Flag* is set to false if the selected cache memory's associativity is 2 (**see Section IV-A**). If the smallest associativity is not 2, *Intersection Flag* is set to true (**see Section IV-A**).

**2.** When *RA* is found in *LT*, CIPARSim selects the cache set sizes one by one, starting from the smallest cache set size, and evaluates all the different cache configurations with the selected cache set size. For a selected cache set size, associativities are selected for evaluation one by one starting from the smallest associativity. When a cache miss occurs in a cache memory, CIPARSim records a cache miss and places a pointer in the selected configuration to point to the location of *RA* in *LT*. On a cache hit, CIPARSim just records a cache hit and continues evaluation to the next cache memory. However, some extra steps are necessary when the selected cache configuration has the smallest associativity. If the smallest associativity is 2 and *RA* is missing in the cache with the smallest associativity, *Track Flag* for that cache is set to false and *RA* is set as the *MRI* of that cache. If *RA* is found in that cache, cache hit is recorded for all the remaining configurations if the *Track Flag* is found true (**see the third intersection property of Section III-A**); and after that evaluation is stopped for *RA*. If *RA* was found in the selected cache but the *Track Flag* is not set to true, the *MRI* entry is checked. If *RA* is found as the *MRI*, cache hit is recorded for all the remaining cache configurations (**see the second intersection property of Section III-A**), *Track Flag* is set to true; and after that evaluation is stopped for *RA*. However, if *RA* is not the *MRI*, a cache hit is recorded for the selected cache and next associativity is picked for evaluation after setting *Track Flag* true. When the smallest associativity is larger than two, and *RA* is found in the selected cache with the smallest associativity, cache hit is declared for all the configurations with the same set size if the *Intersection Flag* is true (**see the first intersection property of Section III-A**). If *RA* is in the smallest associativity cache but the *Intersection Flag* is False, CIPARSim records cache hit for the current cache and continue evaluation to the next configuration. If *RA* was not found in the selected cache with the smallest associativity, CIPARSim updates the *Intersection Flag* of the selected cache according to **the first intersection property of Section III-A**. That means, *Intersection Flag* is set to true if all other cache configurations with the same set size has *RA* and in those larger configurations, *RA* is $((2 \times A_X) - 3)$ (where $A_X$ is the smallest associativity) elements away from the replacement pointers in the direction of replacement. Otherwise, *Intersection Flag* is set to false.

### IV. EXPERIMENTAL PROCEDURE AND RESULTS

To determine the acceleration gained by CIPARSim, we have compared its simulation time with the available single-pass FIFO cache simulators SCUD [14] and DEW [13]. For this purpose, we have re-implemented both SCUD and DEW following the specifications provided by the reference articles. Like CIPARSim, SCUD can simulate FIFO caches with varying set sizes and associativities in a single-pass. However, DEW can simulate FIFO caches with varying set sizes only in a single-pass. As there was no parallelization in use, DEW was repeated multiple times on the simulation machine to simulate different associativities. In each repetition, the trace file was read once in DEW.

To compare the performance of these simulators, twenty-six SPEC CPU2000 benchmark applications (which are mainly general purpose/scientific computation applications) and six Mediabench applications (which are mainly embedded system applications) were used. Applications were executed in "SimpleScalar/PISA 3.0d" [6] to generate the memory trace files. We use SimPoint [8] to identify the most relevant stage in the SPEC CPU2000 programs. Each SPEC CPU2000 application trace is then generated by simulating 300 million instructions within the point identified by SimPoint. This reduction was performed due to the large size of memory traces generated by each SPEC CPU2000 application (SPEC CPU2000 programs ran upwards of 10 billion instructions). Application traces were fed into all the simulators we have implemented. All of these simulators were executed on a machine with a dual core Opteron64 2GHz processor, 8GB of main memory and 1MBytes L2 cache pre-distributed among the processing cores. Note that trace driven cache simulators are used to find the number of cache misses for an application trace and they do not produce wrong/different results even if executed on a general purpose processor or embedded processor, as long as the trace file is same. Due to space limitation, simulation results are presented only for six SPEC CPU2000 applications and five Mediabench applications in this article. Name of the applications are presented in the first column of Table II. The applications are selected depending on their total number of memory accesses. We have selected some applications with very few memory accesses (e.g.; sixTrack and JPEG decode), some applications with many memory accesses (e.g.; eon and MPEG2 decode) and the remaining applications in-between the extremes (e.g.; ammp and G721 decode). The number of memory accesses in each application is presented in the sixth column in Table II.

| Cache Set Size=$2^i$ | $0 <= i <= 14$ |
| Line Size=$2^i$ Bytes | $2 <= i <= 6$ |
| Associativity=$2^i$ | $1 <= i <= 4$ |

TABLE I
CACHE CONFIGURATION PARAMETERS

To compare CIPARSim's performance with SCUD and DEW, 300 FIFO cache configurations (non unified caches) were simulated on each of the three simulators to generate each application's total number of cache misses. Table I shows how the 300 FIFO cache configurations were derived from the cache parameters.

In Table II, the simulation times of DEW, SCUD and CIPARSim have been been presented. Column 2 presents the cache line size (Only 4, 16 and 64 Bytes are presented due to space limitation). Columns 3 to 5 present the simulation time for DEW, SCUD and CIPARSim respectively to simulate the 60 cache configurations of Table I for the particular cache line size. For each and every application, CIPARSim showed significantly faster performance than DEW and SCUD. Over DEW, CIPARSim showed the highest speedup of 10 times for application "eon" and block size 64 Bytes. In this case, DEW's simulation time was 53.35min and CIPARSim's time was 5.16min. Over SCUD, CIPARSim showed 5 times speedup at best for application "eon" and cache line size 4 Bytes. In this case, CIPARSim's simulation time was 27.31min whereas SCUD's execution time was 1.19hour. On average, CIPARSim is 5 times faster than DEW and 3 times faster than SCUD. CIPARSim's speedup (which is (*DEW or SCUD simulation time*)/(*CIPARSim's time*)) has been presented in Figure 3 for all the six SPEC CPU2000 and five Mediabench applications.

| Application | Cache Line Size (Byte) | Simulation time h=Hour, m=Min, s=Sec | | | Total Accesses Per Cache (Million) | Total Hits (Million) | | Hits Predicted by (Million) | | |
|---|---|---|---|---|---|---|---|---|---|---|
| | | DEW | SCUD | CIPARSim | | $A \geq 2$ | $A \geq 4$ | Intersection1 | Intersection2 | Intersection3 |
| SPEC CPU2000 (General purpose/Scientific computation applications) | | | | | | | | | | |
| sixTrack | 4 | 1.29h | 55.07m | 20.19m | 378.43 | 13858.83 | 10939.62 | 4552.53 | 9105.06 | 73.70 |
| swim | 4 | 1.53h | 1.32h | 30.19m | 400.19 | 12726.93 | 10033.50 | 4210.85 | 8421.70 | 2.10 |
| ammp | 4 | 1.31h | 1.46h | 20.76m | 430.11 | 17420.09 | 13657.97 | 5911.58 | 11823.15 | 300.09 |
| mesa | 4 | 1.51h | 1.77h | 22.94m | 431.83 | 25691.41 | 12212.40 | 4659.59 | 9319.19 | 218.31 |
| eon | 4 | 1.76h | 1.19h | 27.31m | 519.19 | 30756.26 | 15576.38 | 6124.25 | 12248.51 | 395.40 |
| mcf | 4 | 1.68h | 1.18h | 31.57m | 561.81 | 33567.48 | 19166.40 | 9072.49 | 18144.98 | 141.20 |
| sixTrack | 16 | 46.64m | 18.29m | 7.43m | 378.43 | 20764.40 | 15771.44 | 8943.24 | 17886.48 | 90.30 |
| swim | 16 | 55.13m | 30.91m | 13.05m | 400.19 | 20363.77 | 15466.99 | 8795.02 | 17590.04 | 2.78 |
| ammp | 16 | 50.19m | 30.06m | 8.12m | 430.11 | 23711.23 | 18002.26 | 10075.06 | 20150.12 | 316.29 |
| mesa | 16 | 53.28m | 34.10m | 8.73m | 431.83 | 25746.93 | 17729.76 | 9849.95 | 19699.91 | 162.79 |
| eon | 16 | 1.09h | 24.91m | 10.23m | 519.19 | 30941.72 | 19472.70 | 10786.72 | 21573.44 | 209.95 |
| mcf | 16 | 1.26h | 43.92m | 17.07m | 561.81 | 33550.93 | 22284.74 | 12719.55 | 25439.10 | 157.75 |
| sixTrack | 64 | 36.29m | 10.58m | 4.31m | 378.43 | 22226.71 | 16771.51 | 9893.02 | 19786.04 | 178.03 |
| swim | 64 | 43.40m | 15.81m | 5.73m | 400.19 | 22933.43 | 17319.52 | 10112.44 | 20224.87 | 59.67 |
| ammp | 64 | 42.48m | 15.31m | 5.68m | 430.11 | 25173.83 | 18973.82 | 11033.15 | 22066.30 | 380.59 |
| mesa | 64 | 43.64m | 17.37m | 6.01m | 431.83 | 25662.26 | 18897.92 | 10996.83 | 21993.65 | 247.45 |
| eon | 64 | 53.35m | 13.51m | 5.16m | 519.19 | 30748.23 | 22883.84 | 13228.38 | 26456.76 | 403.44 |
| mcf | 64 | 1.06h | 27.35m | 9.65m | 561.81 | 33548.54 | 23715.15 | 13782.77 | 27565.54 | 160.14 |
| Mediabench (Embedded system applications) | | | | | | | | | | |
| JPEG Dec | 4 | 55.40s | 1m | 21.11s | 7.62 | 370.14 | 287.37 | 130.08 | 260.17 | 13.14 |
| G721 Dec | 4 | 20.01m | 21.50m | 7.21m | 154.86 | 7287.86 | 5622.24 | 2729.84 | 5459.69 | 50.83 |
| MPEG2 Dec | 4 | 3.39h | 4.28h | 1.21h | 1411.43 | 59820.72 | 47031.11 | 18475.22 | 36950.44 | 1267.91 |
| JPEG Enc | 4 | 2.94m | 3.01m | 1.04m | 25.68 | 1293.63 | 1001.89 | 463.27 | 926.55 | 51.71 |
| G721 Enc | 4 | 19.45m | 21.22m | 5.83m | 155.00 | 7322.09 | 5642.86 | 2757.57 | 5515.13 | 42.31 |
| JPEG Dec | 16 | 42.68s | 28.18s | 11.51s | 7.62 | 427.59 | 326.36 | 168.83 | 337.65 | 14.92 |
| G721 Dec | 16 | 14.54m | 9.86m | 4.35m | 154.86 | 8541.49 | 6488.32 | 3599.14 | 7198.28 | 77.98 |
| MPEG2 Dec | 16 | 2.35h | 1.91h | 39.52m | 1411.43 | 75642.60 | 57897.91 | 29915.88 | 59831.76 | 1028.83 |
| JPEG Enc | 16 | 2.24m | 1.42m | 42.94s | 25.68 | 1462.68 | 1115.48 | 580.93 | 1161.86 | 56.45 |
| G721 Enc | 16 | 13.89m | 9.64m | 3.95m | 155.00 | 8564.19 | 6503.45 | 3657.84 | 7315.68 | 18.13 |
| JPEG Dec | 64 | 35.97s | 15.29s | 9.43s | 7.62 | 447.59 | 338.52 | 184.64 | 369.28 | 15.91 |
| G721 Dec | 64 | 12.01m | 4.71m | 2.51m | 154.86 | 9084.08 | 6852.10 | 3922.17 | 7844.35 | 224.01 |
| MPEG2 Dec | 64 | 1.89h | 53.05m | 24.01m | 1411.43 | 82069.41 | 62078.42 | 34777.99 | 69555.98 | 1405.96 |
| JPEG Enc | 64 | 2m | 52.59s | 27.05s | 25.68 | 1512.13 | 1144.47 | 629.99 | 1259.99 | 45.84 |
| G721 Enc | 64 | 11.45m | 4.28m | 2.53m | 155.00 | 9126.34 | 6876.91 | 4015.23 | 8030.47 | 82.12 |

TABLE II
SIMULATION TIME AND PERFORMANCE ANALYSIS

During simulation of CIPARSim, we have recorded the number of cache hits predicted by the intersection properties discussed in this paper. Columns 10 and 11 present the total number of cache hits predicted by the second and third intersection property respectively while simulating the 60 cache configurations for each cache line size. The total number of cache hits that occurred is presented in column 7. From the results, it can be seen that the second and third intersection properties together can predict up to 90% of the total cache hits (for "sixTrack" the total number of hits is 22 billion with 19 billion hits predicted by the second intersection property and 178 million are predicted by the third intersection property).

To check the effectiveness of the first intersection property, we have added the *Intersection Flags* with the tags in associativity 4. As the first intersection property can be applied when the smallest associativity is greater than 2, we have presented, in column 8, the total number of cache hits that would occur when associativities 4, 8 and 16 were considered only. From the results, it can be seen that the first intersection property alone can predict up to 60% of the total hits which is observed again for "sixTrack" and block size 64 Bytes. In this case, the total number of cache hits is 17 billion, 10 billion of which are predicted by the first intersection property.

As a very large number of cache hits were predicted (on average, 65% hits are predicted) in CIPARSim by the intersection properties, much of the time consuming simulation steps were avoided. In addition to the profound role of intersection properties in reducing simulation time in CIPARSim, the data structure also played a noticeable role. Unlike SCUD, CIPARSim divides the look-up table into smaller sets. Therefore, binary search needs to search a small set of elements to find the requested memory block quickly. Once the memory block is found in the lookup table, fast bit operations are performed to determine cache hits and misses. Bit arrays not only helped to reduce simulation time, they also made CIPARSim space generous. Like SCUD, CIPARSim uses look-up table and simulation tree; however, CIPARSim's space consumption is almost 55% less than SCUD as bit arrays are used in look-up table entries. DEW consumes much less space than SCUD and CIPARSim; however, its space generous data structure does not allow it to simulate cache configurations with varying set sizes and associativities together.

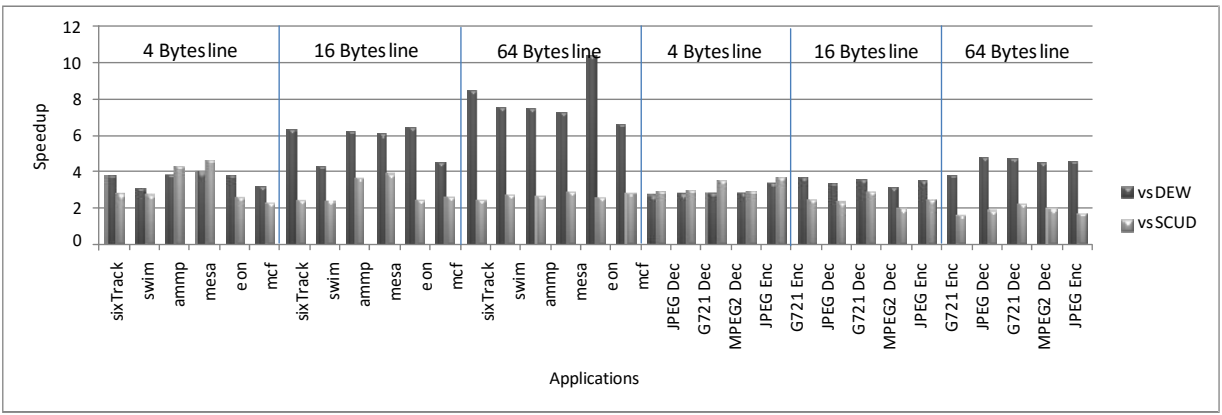

Fig. 3. Speedup in CIPARSim

## V. CONCLUSION

To assist in the single-pass simulation of FIFO caches, a new kind of cache property called "Intersection property" has been introduced in this paper. By predicting a large number of cache hits quickly and accurately, intersection properties can help to avoid simulation of large number of caches during single-pass simulation. In this paper, three cache intersection properties have also been discussed that can be used in single-pass simulation. Utilizing the intersection properties, the single-pass FIFO cache simulator, "CIPARSim", presented in this paper shows greatly improved performance compared to the available simulators in finding the cache miss rate of an application on various FIFO cache configurations.